\documentclass{elsart1p}

\usepackage{graphics}
 \usepackage{graphicx}
\usepackage{epsfig}

\usepackage{amssymb}

\begin{document}

\begin{frontmatter}



\title{Spectroscopy of Mesons with Heavy Quarks}


\author{Shi-Lin Zhu}

\address{Department of Physics, Peking University, Beijing 100871, China}
\ead{zhusl@phy.pku.edu.cn}

\ead[url]{http://inpc2007.riken.jp/P/P5-zhu.pdf}

\begin{abstract}
I will give a concise overview of mesons with heavy quarks
including p-wave charmed mesons and charmonium (or
charmonium-like) states such as X(3872), Y(4260), X(3940),
Y(3940), Z(3930) etc. The effect from the nearby S-wave open
channels on the quark model spectrum is emphasized.
\end{abstract}

\end{frontmatter}

%%%%%%%%%%%%%%%%%%%%%%%%%%%%%%%%%%%
\section{QCD and hadron physics}\label{sec1}
%%%%%%%%%%%%%%%%%%%%%%%%%%%%%%%%%%%

QCD is the underlying theory of strong interaction, which has
three fundamental properties: asymptotic freedom, confinement, and
approximate chiral symmetry and its spontaneous breaking.
Perturbative QCD has been tested to very high accuracy. But the
low energy sector of QCD (i.e., hadron physics) still remains very
challenging. Precision-test of Standard Model and search for new
physics require good knowledge of hadrons as inputs such as parton
distribution functions, hadron distribution amplitudes etc.

The motion and interaction of hadrons differ from those of nuclei
and elementary particles like quarks, gluons, leptons and gauge
bosons. Hadron physics is the bridge between nuclear physics and
particle physics. The famous Higgs mechanism contributes around 20
MeV to the nucleon mass through current quark mass. Nearly all the
mass of the visible matter in our universe comes from the
non-perturbative QCD interaction. Therefore study of hadron
spectroscopy explores the mechanism of confinement and chiral
symmetry breaking, and the mass origin.

Quark model is quite successful in the classification of hadrons
although it's not derived from QCD. Any state with quark content
other than $q\bar q$ or $qqq$ is beyond the naive quark model. But
quark model can't be the whole story. QCD may allow much richer
hadron spectrum such as glueballs, hybrid mesons/baryons,
multiquark states, hadron molecules. Although experimental search
of these non-conventional states started many years ago, none of
them has been established without controversy experimentally!

Typical signatures of these non-conventional states include:
\begin{itemize}

\item Exotic flavor quantum number like $\theta^+$

\item Exotic $J^{PC}$ quantum number like $1^{-+}$ exotic meson

\item Overpopulation of the QM spectrum like the scalar isoscalar
spectrum below 1.9 GeV: $\sigma$, $f_0(980)$, $f_0(1370)$,
$f_0(1500)$, $f_0(1710)$, $f_0(1790)$, $f_0(1810)$.

\end{itemize}

In this talk I will review the recent progress on the P-wave
charmed mesons in the past several years, especially
$D_0^*(2308/2407)$, $D_1^*(2427)$, $D_{s0}^*(2317)$,
$D_{s1}^*(2460)$. I will also discuss the important progress on
the charmonium and charmonium-like system including X(3872),
X(3940), Y(3940), Y(4260), Z(3930) etc. Interested readers may
consult the recent review on new hadron states \cite{zhu-ijmpa}.

%%%%%%%%%%%%%%%%%%%%%%%%%%%%%%%%%%%
\section{Charmed mesons}\label{sec2}
%%%%%%%%%%%%%%%%%%%%%%%%%%%%%%%%%%%

Heavy quark expansion provides a systematic method to deal with
hadrons with a single heavy quark. The angular momentum $j_l$ of
the light quark in the $Q\bar q$ system is a good quantum number
in the heavy quark limit. Heavy mesons form doublets with
different $j_l$ and parity. For $L=0$, we have the ground doublet
$(0^-, 1^-)$. For the $L=1$ P-wave states, we have two doublets:
$(0^+,1^+)$, $(1^+, 2^+)$. The $(0^-, 1^-)$ and $(1^+, 2^+)$
doublets agree with theoretical expectation quite well. In
contrast, there are two puzzles with the $(0^+,1^+)$ doublet. The
heavy-light system is the QCD "hydrogen"!

The non-strange $(0^+,1^+)$ doublet decay through s-wave. They are
very broad with a width around 300 MeV. The non-strange $(1^+,
2^+)$ doublet decay through d-wave. They are narrow with a width
around 20 MeV. There was one measurement of the $1^+$ mass from
Belle Collaboration $m_{D^\ast_{1}}=2427\pm 26\pm 25$ MeV with a
large width $384^{+107}_{-75}\pm 74$ MeV \cite{focus1}, where we
use the star to indicate this $1^+$ state belongs to the
$(0^+,1^+)$ doublet. For the $0^+$ state, there were two
measurements. FOCUS Collaborations reported $m_{D^\ast_{0}}=2308
\pm 17 \pm 32$ MeV with a width $276\pm 21\pm 63 $ MeV
\cite{belle1} while BELLE observed it at $2407\pm 21\pm 35$ MeV
with a width $240\pm 55 \pm 59$ MeV \cite{focus1}.

For the strange doublet, we have $m_{D^\ast_{s0}}=2317$ MeV,
$m_{D^\ast_{s1}}=2459$ MeV \cite{pdg}. $D_{s0}^*$ and $D_{s1}^*$
lie below DK (D*K) threshold. They are roughly 160 MeV below quark
model prediction \cite{qm}. Both of them are extremely narrow.
Their strong decays violate isospin symmetry and occur with help
of a virtual $\eta$ meson: $D_{s0}^*\to D_s \eta \to D_s \pi^0$.
The mass of $D_{s0}^\ast$ from three lattice QCD simulations is
still larger than experimental value \cite{bali2,dougall,soni}.
Naively one would expect that $D_{s0}^*(2317)$ lies 100 MeV above
$D_0^*(2308/2407)$ because of the mass difference between strange
and up quarks. Now arise the two puzzles: (1) why is the mass of
$D_{s0}^* (D_{s1}^*)$ so low? (2) why are $D_{s0}^*$ and $D_0^*$
nearly degenerate?

The low mass of $D_{s0}^* (D_{s1}^*)$ inspired various tetraquark
schemes. For example, if $D_0^*$ and $D_{s0}^*$ were in the
anti-symmetric $\bar 3$ multiplet, their flavor wave functions are
\cite{pol}
\begin{equation}
|D_0^\ast\rangle ={1\over 2} |c(s(\bar u \bar s -\bar s\bar u)- d
(\bar d\bar u -\bar u \bar d))\rangle \; .
\end{equation}
\begin{equation}
|D_{s0}^\ast\rangle ={1\over 2} |c(u(\bar u \bar s -\bar s\bar u)-
d (\bar d\bar s -\bar s \bar d))\rangle \; .
\end{equation}
Since they contain the same amount of strange, they would have
roughly the same mass.

But tetraquarks always contain the color-singlet times
color-singlet component in their color wave function. They would
fall apart easily and become very broad. There always exist two
difficult issues for the tetraquark interpretation: (1) where are
the conventional $(0^+,1^+)$ states in the quark model? (2) where
are those partner states in the same tetraquark multiplet? In
fact, Babar collaboration scanned around 2.31 GeV, 2.46 GeV and
below 2.7 GeV. Not surprisingly, they found neither additional
$(0^+,1^+)$ states nor their spin-flavor partner states.

Belle, Babar and Cleo collaborations measured the ratio of
radiative and strong decay widths of $D_{s0}^*$ and $D_{s1}^*$,
which is collected in Table \ref{tab-ratio}.
\begin{table}[h]
\caption{Comparison between experimental ratio of $D_{sJ}(2317,
2460)$ radiative and strong decay widths and theoretical
predictions from light-cone QCD sum rule
approach.\label{tab-ratio}}
\begin{center}
\begin{tabular}{cccc|c}
\hline & Belle & Babar   & CLEO  & LCQSR\\ \hline $\frac{\Gamma
\left( D^*_{sJ}(2317) \rightarrow D_{s}^{\ast }\gamma \right) }{
\Gamma \left( D^*_{sJ}(2317)\rightarrow D_{s}\pi ^{0}\right) }$ &
$<0.18$ \cite{belle2}& & $<0.059$& 0.13 \\
\hline $\frac{\Gamma \left(D_{sJ}(2460) \rightarrow D_{s}\gamma
\right) }{ \Gamma \left( D_{sJ}(2460)\rightarrow D_{s}^{\ast }\pi
^{0}\right) }$ & $0.55\pm0.13$  & $0.375\pm0.054$
&$<0.49$ & 0.56\\
& $\pm0.08$ \cite{belle2} & $\pm0.057$ \cite{babar5} & &\\
\hline $\frac{\Gamma \left( D_{sJ}(2460)\rightarrow D_{s}^{\ast
}\gamma \right) }{\Gamma \left( D_{sJ}(2460)\rightarrow
D_{s}^{\ast }\pi ^{0}\right) }$
& $<0.31$ \cite{belle2}& &$<0.16$ & 0.02 \\
\hline $\frac{\Gamma \left( D_{sJ}(2460)\rightarrow
D^*_{sJ}(2317)\gamma \right) }{\Gamma \left(
D_{sJ}(2460)\rightarrow
D_{s}^*\pi^0 \right) }$  &  & $ < 0.23$   \cite{babar4}& $ < 0.58$ & 0.015 \\
\hline
\end{tabular}
\end{center}
\end{table}
Assuming $D_{s0}^*$ and $D_{s1}^*$ are conventional $c\bar s$
mesons, theoretical ratio from light-cone QCD sum rules
\cite{col1,wei1} and $^3P_0$ model \cite{lu} is consistent with
Belle/Babar's recent data.

Coupled channel effects may be the origin of the low mass puzzle
of $D_{s0}^*$ and $D_{s1}^*$ since they have the same quantum
number as S-wave $DK (D^*K)$ continuum and lie very close to $DK
(D^*K)$ threshold (within 46 MeV). Moreover the $D_{s0}^*DK$
coupling is very large. Within the quark model, the configuration
mixing effects between the "bare" $(0^+, 1^+)$ and $DK (D^*K)$ may
lower the mass of $D_{s0}^*$ and $D_{s1}^*$. Within the QCD sum
rule framework, the DK continuum contribution may be important
\cite{dai1}. This mechanism also provides a possible explanation
why quenched lattice QCD simulations get a higher mass since
quenched approximation ignores the meson loop.

%%%%%%%%%%%%%%%%%%%%%%%%%%%%%%%%%%%
\section{Charmonium or charmonium-like states}\label{sec3}
%%%%%%%%%%%%%%%%%%%%%%%%%%%%%%%%%%%

The charmonium system is the playground of new phenomenological
models of the low-energy strong interaction since QCD can not be
solved analytically at present. The potential model is widely
used. Usually there are three pieces in the potential. The first
one is a central potential from one gluon exchange and the linear
confinement. The second term is the spin-spin interaction which
splits the spin singlet and triplet states like $J/\psi$ and
$\eta_c$. The third piece is the spin-orbit interaction which is
responsible for the splitting among states like $\chi_{c0,1,2}$.

There has been important progress in the charmonium spectroscopy
in the past few years. Several previously "missing" states were
observed, which are expected in the quark model. Quite a few
unexpected states are discovered experimentally, seriously
challenging the quark model. These new states were named
alphabetically as XYZ etc. Aspects of these XYZ states have been
reviewed in literature, for example in Refs.
\cite{review8a,review8b,review8c,review8d,review8e,review8f}.

\subsection{Z(3930)}

Belle collaboration observed Z(3930) in the $D\bar D$ channel in
the electron positron annihilation \cite{belle9z}. Since this
state was produced through the two photon reaction, its parity and
C-parity are even. From angular distribution of final states, its
angular momentum was found to be two. Its total width is around 20
MeV. The property of this tensor state matches well with
$\chi_{c2}^\prime$ in the quark model, although its expected
$D^\ast \bar D$ mode has not been discovered yet.

It's interesting to compare Z(3930) with quark model prediction of
the mass of $\chi_{c2}^\prime$, which ranges from 3972 MeV to 4030
MeV. In other words, quark model predictions of the
$\chi_{c2}^\prime$ mass is always 40-100 MeV higher. I want to
emphasize that this may be the typical accuracy of quark model for
the higher charmonium states above open charm decay threshold.

\subsection{X(3940)}

In the recoil mass spectrum of $J/\psi$ in the electron positron
annihilation, Belle observed X(3940) in the $\bar D D^\ast$
channel but not in the $D\bar D$ and $\omega J/\psi$ modes
\cite{belle9x}. Its C-parity is even with a width less than 52
MeV. Such a decay pattern is typical of $\chi_{c1}^\prime$.

But the ground state $\chi_{c1}$ is not seen in the same
experiment. Hence X(3940) does not look like $\chi_{c1}^\prime$.
Instead X(3940) may be $\eta_c^{\prime\prime}$ except that it's
100 MeV below the QM prediction.

\subsection{Y(3940)}

Belle collaboration observed a broad threshold enhancement Y(3940)
in $\omega J/\psi$ channel in the $B\to K \omega J/\psi$ decay
\cite{belle9y}. If this enhancement is taken as a particle, its
width is around 92 MeV. The hidden charm decay mode $Y(3940)\to
\omega J/\psi$ violates $SU_F(3)$ flavor symmetry. It's very
unusual its width is larger than 7 MeV! Such a decay pattern is
very puzzling while its dominant decay mode remains to be
discovered. This state has not been confirmed by other
collaborations yet.

\subsection{X(3872)}

\subsubsection{Experimental information and its quantum number}

Belle collaboration first observed X(3872) in the $\pi^+\pi^-
J/\psi$ channel in the $B\to K\pi^+\pi^- J/\psi$ decays
\cite{belle8a}. The di-pion spectrum looks like a rho meson. In
the same experiment, a sharp $\psi^\prime$ signal was also
observed.

Later it was also observed in the $\pi^+\pi^-\pi^0 J/\psi$ mode
\cite{belle8b}. The three pion spectrum peaks around a virtual
omega meson. According to PDG \cite{pdg}, its mass is $3871.2\pm
0.54$ MeV and width less than 2.3 MeV, which is the typical
detector resolution. It's important to note that the $\rho J/\psi$
decay mode violates isospin symmetry!

Both CDF and D0 collaborations confirmed X(3872) in the
$\pi^+\pi^- J/\psi$ channel in the proton anti-proton collision
\cite{cdf8a,d08}. Again the di-pion spectrum looks like a rho
meson \cite{cdf8b} and a very clear $\psi^\prime$ signal was
observed in the same experiments. In other words, the production
properties of X(3872) are very similar to those of $\psi^\prime$,
which is a pure charmonium state.

Both Babar and Belle collaboration observed the radiative decay
mode $X(3872)\to \gamma J/\psi$ \cite{belle8b,babar8c}. Therefore
the C-parity of X(3872) is even. From angular correlations of
final states, Belle collaboration ruled out the $0^{++}$ and
$0^{-+}$ possibilities and favors the $1^{++}$ assignment
\cite{belle8c}. The analysis of CDF collaborations allows only
$1^{++}$ and $2^{-+}$ \cite{cdf8c}. Hence the quantum number of
X(3872) is probably $1^{++}$. But the $2^{-+}$ possibility is not
ruled out by experiments.

There are theoretical arguments against the $2^{-+}$ possibility
in the non-relativistic quark model \cite{zhaogd}. Since the
$2^{-+}$ charmonium is the spin-singlet D-wave state and $J/\psi$
is the spin-triplet S-wave state, E1 transition $2^{-+}\to
J/\psi\gamma $ is forbidden in the non-relativistic limit. On the
other hand, the D-wave radial wave function is orthogonal to the
S-wave radial wave function, therefore M1 transition $2^{-+}\to
J/\psi\gamma $ is also forbidden. Belle and BaBar collaborations
observed the radiative decay mode. Therefore X(3872) is unlikely
to be the $2^{-+}$ charmonium. But will relativistic corrections
change this picture?

\subsubsection{Is X(3872) a molecular state?}

X(3872) sits exactly on the ${\bar D}^0 D^{0\ast}$ threshold and
lies very close to the $\rho J/\psi$, $\omega J/\psi$ and ${\bar
D}^+ D^{-\ast}$ thresholds. It is extremely narrow and around 100
MeV below quark model prediction of $\chi_{c1}^\prime$. Its hidden
charm modes are quite important while the $\rho J/\psi$ decay mode
violates isospin symmetry. All the above facts stimulated several
groups to propose X(3872) could be a molecular state
\cite{close8,v8,wong8,swanson8,tornqvist8}.

Especially Swanson proposed \cite{swanson8} that X(3872) is mainly
a ${\bar D}^0 D^{0\ast}$ molecule bound by both quark and pion
exchange. Its wave function also contains small but important
$\rho J/\psi$, $\omega J/\psi$ and ${\bar D}^+ D^{-\ast}$
components. The molecule picture explains the proximity to the
${\bar D}^0 D^{0\ast}$ threshold and hidden charm decay modes
quite naturally. This model has been very popular.

But experimental evidence against the molecular assignment is
accumulating. The radiative decay mode is clean and ideal to test
the model. Two experiments measured this ratio. The value from
Belle collaboration is \cite{belle8b}
\begin{equation}
{B\left(  X(3872)\to \gamma J/\psi \right)\over B \left(
X(3872)\to \pi^+\pi^- J/\psi \right) } = 0.14\pm 0.05
\end{equation}
and that from Babar collaboration is \cite{babar8c}
\begin{equation}
{B\left(  X(3872)\to \gamma J/\psi \right)\over B \left(
X(3872)\to \pi^+\pi^- J/\psi \right) } \approx 0.25
\end{equation}
while the theoretical prediction from the molecular model is
0.007.

Belle collaboration reported a near-threshold enhancement in the
$D^0\bar D^0\pi^0$ system with a mass $3875.4\pm 0.7
^{+1.2}_{-2.0}$ MeV \cite{belle8d}. From this measurement
\begin{equation}
{B\left( X(3872)\to D^0\bar D^0\pi^0 \right)\over B\left(
X(3872)\to \pi^+\pi^- J/\psi \right) } = 9.4^{+3.6}_{-4.3}
\end{equation}
while the theoretical prediction is 0.054. From Table I in Ref.
\cite{belle8d}, we have
\begin{equation}
{B\left( B^0\to X(3872)K^0 \right)\over B \left( B^+\to X(3872)K^+
\right) } \approx 1.62
\end{equation}
while the theoretical prediction is less than 0.1.

Very recently Babar collaboration observed X(3872) in the ${\bar
D}^0 D^{0\ast}$ invariant mass spectrum with a mass $3875.6\pm 0.7
^{+1.4}_{-1.5}$ MeV \cite{babar-talk}, which agrees with the value
from Ref. \cite{belle8d} very well. Such a value is clearly above
the ${\bar D}^0 D^{0\ast}$ threshold and does not support the
molecular picture.

\subsubsection{Could X(3872) still be a $1^{++}$ charmonium?}

Recall that the production properties of X(3872) are similar to
those of $\psi^\prime$ and the typical quark model accuracy is
around 100 MeV for charmonium states above open-charm decay
threshold. Deviation around 100 MeV from quark model prediction
may be still acceptable. Very interestingly, a recent lattice
simulation by CLQCD collaboration claimed that $\chi_{c1}^\prime$
lies around 3853 MeV \cite{liuchuan}. The $1^{++}$ charmonium
assignment really deserves serious attention
\cite{chao8,suzuki8,zhu-ijmpa}!

There are three main obstacles of $1^{++}$ charmonium assignment.
However, possible solutions exist which I list below:
\begin{itemize}
\item Low mass

Strong S-wave coupled channel effects may lower its mass?

\item Large isospin symmetry breaking $\rho J/\psi$ decay

Hidden charm decay can happen through rescattering mechanism
\cite{fsi8,meng8}: $ X \to {\bar D}^0 D^{0\ast} + {\bar D}^+
D^{-\ast}\to \rho J/\psi (\omega J/\psi)$. There is isospin
symmetry breaking in the mass of $\bar D D^\ast$ pair since
$D^+(D^{-\ast})$ is heavier than $D^0(D^{0\ast})$. The $\rho
J/\psi$ mode has much larger phase space than $\omega J/\psi$ mode
since the rho meson is very broad. All the above factors may
combine to make a sizable $\rho J/\psi$ decay width.

\item Extremely narrow width

The total width of X(3875) needs some exotic schemes such as
decreasing quark pair creation strength of $^3P_0$ model near
threshold \cite{chao8a}.

\end{itemize}

%%%%%%%%%%%%%%%%%%%%%%%%%%%%%%%%%%%%%%%%%
\section{Y(4260)}\label{sec4}
%%%%%%%%%%%%%%%%%%%%%%%%%%%%%%%%%%%%%%%%%

BABAR collaboration observed a charmonium state around 4.26 GeV in
the $\pi^+\pi^- J/\psi$ channel \cite{babar7}. Since this
resonance is observed in the $e^+e^-$ annihilation through initial
state radiation (ISR), its spin-parity is known $J^{PC}=1^{--}$.
Later several other collaborations confirmed Y(4260)
\cite{cleo7,cleo7a,belle7,babar7c}. The cental values of its mass
and width from various measurement ar collected in Table
\ref{tab-y4260}.
\begin{table}[h]
\begin{center}
\begin{tabular}{|c|c|c|c|c|}\hline
&Babar&CLEO-c&CLEO III&Belle\\
Events&125&50&14&165\\
Mass&4259&4260&4283&4295\\
Width&88&&70&133\\ \hline
\end{tabular}
\end{center}
\caption{\baselineskip 15pt The central values of the extracted
mass and width of Y(4260) from various experimental
measurements.\label{tab-y4260}}
\end{table}

However Y(4260) was not seen in the $e^+e^-$ annihilation. In
fact, R distribution dips around 4.26 GeV. Its leptonic width is
small: $\Gamma(Y\to e^+e^-)<240$ eV \cite{mo} and its hidden charm
decay width is large: $\Gamma (Y\to J/\psi \pi \pi)>1.8$ MeV!

According to the PDG assignment of the $1^{--}$ charmonium, there
are four S-wave states $J/\psi$, $\psi(3686)$, $\psi(4040)$,
$\psi(4415)$ and two D-wave states $\psi(3770)$, $\psi(4160)$.
Naively one would expect the $3^3D_1$ state lying above 4.4 GeV.
In other words, there is no suitable place for Y(4260) in the
quark model spectrum. All the above states have a sharp peak in R
distribution. But Y(4260) has a dip! The discovery of Y(4260)
indicates the overpopulation of the $1^{--}$ spectrum if PDG
classification of the observed $1^{--}$ charmonium is correct.

From BES and CLEOc's recent measurement, the hidden charm decay
width of $\psi^{\prime\prime}$: $\Gamma(\psi^{\prime\prime}\to
J/\psi \pi\pi)\approx 50$ keV \cite{pdg}. If Y(4260) is a
charmonium state, one might expect a comparable $J/\psi \pi\pi$
decay width instead of $\Gamma (Y\to J/\psi \pi\pi)>1.8$ MeV.
Similar di-pion transitions from $\psi(4040)$ or $\psi(4160)$ were
not observed in the same experiments. One may wonder whether the
conventional charmonium assignment is in trouble.

Virtual photon does not couple to glues directly and glueballs
easily decay into light hadrons which were not observed
experimentally. So Y(4260) does not look like a glueball.

Although it lies close to $\bar D D_1(2420)$, $\bar D D_1^\ast$ or
${\bar D}_0^\ast(2310) D^\ast$ thresholds, Y(4260) does not seem
to arise from the threshold or coupled-channel effects since the
$J/\psi \pi\pi$ spectrum is very symmetric. There is no obvious
distortion from nearby thresholds.

Could Y(4260) be a tetraquark? Tetraquark falls apart into $D\bar
D$ very easily. So $D\bar D$ should be one of the dominant decay
modes. Its width would be much larger than 90 MeV! Moreover, if
the isoscalar component of the photon had produced Y(4260) with
$I^G=0^-$, its isovector componet would also have produced
$Y^\prime (4260)$ with $I^G=1^+$, which decays into $J/\psi \pi^+
\pi^-\pi^0$. This possibility had been ruled out by Babar
collaboration \cite{zhu7}!

Several groups suggested that Y(4260) may be a a hybrid charmonium
in 2005 \cite{zhu7,kou7,close7}. Its mass leptonic width, total
width, production cross section, decay pattern (hidden charm vs
open charm), flavor blind decays into $J/\psi \pi\pi$ and $J/\psi
K\bar K$, overpopulation of $1^{--}$ spectrum and its large hidden
charm decay width satisfy the very naive expectation of a hybrid
charmonium state.

It's very interesting to recall that Quigg and Rosner predicted
one $1^{--}$ charmonium state at 4233 MeV using the logarithmic
potential thirty years ago, which was identified as the 4S state
\cite{rosner7}. In order to study possible effects of color
screening and large string tension in heavy quarkonium spectra,
Ding, Chao, and Qin also predicted their 4S charmonium state
exactly at 4262 MeV twelve years ago \cite{chao7}! Their potential
is quite simple:
\begin{equation}
V(r)=-{4\over 3}{\alpha_s\over r} + {T\over \mu}(1-e^{-\mu r})
\end{equation}
where $T$ is the string tension and $\mu$ is the screening
parameter. With such a perfect agreement, one may wonder whether
PDG assignment misses one $1^{--}$ charmonium state in the quark
model. Or does the same traditional quark potential hold for
higher states far above strong decay threshold? However, two
serious challenges remain for the conventional quark model
interpretation: (1) how to generate the huge $J/\psi \pi\pi$ decay
width? (2) How to explain the dip in the R distribution?

%%%%%%%%%%%%%%%%%%%%%%%%%%%%%%%%%%%%%%%%%
\section{Summary}\label{sec5}
%%%%%%%%%%%%%%%%%%%%%%%%%%%%%%%%%%%%%%%%%

After four years' extensive theoretical and experimental efforts,
the situation of $D_{sj}$ mesons is almost clear: both
$D_{s0}^\ast (2317)$ and $D_{s1}^\ast(2460)$ are probably $c\bar
s$ states. But the higher charmonium sector is still very
controversial
\begin{itemize}

\item  Z(3930) is very probably $\chi_{c2}^\prime$

\item  X(3940) may be $\eta_c^{\prime\prime}$

\item Y(3940) needs confirmation

\item X(3872) may be a candidate of $\chi_{c1}^\prime$ (or
molecule)

\item Y(4260) may be a candidate of hybrid charmonium (or
charmonium).

\end{itemize}

BESIII in Beijing will start taking data this year and will
increase its database by 100 times. Jlab, B factories and other
facilities are increasing the database continuously. J-PARC will
start running at the end of next year. There will be great
progress in the search of non-conventional hadrons and more
unexpected...

%%%%%%%%%%%%%%%%%%%%%%%%%%%%%%%%%%%%%%%%%
\section*{Acknowledgment}
%%%%%%%%%%%%%%%%%%%%%%%%%%%%%%%%%%%%%%%%%

The author thanks the organizers of the international nuclear
physics conference 2007 for their kind invitation. This project
was supported by the National Natural Science Foundation of China
under Grants 10421503 and 10625521, and Ministry of Education of
China, FANEDD.

\end{document}